\documentclass[aps,prb,superscriptaddress,twocolumn,10pt]{revtex4-1}

\usepackage[english]{babel}
\usepackage[utf8]{inputenc}
\usepackage{xcolor}
\usepackage{tabularx}
\usepackage{amssymb}
\usepackage{amsmath}
\usepackage{graphicx}
\usepackage{float} 
\usepackage{bm}
\usepackage{booktabs}
\usepackage{adjustbox}
\usepackage{amsmath}

\usepackage{silence}
\makeatletter
\newcommand{\mathleft}{\@fleqntrue\@mathmargin0pt}
\newcommand{\mathcenter}{\@fleqnfalse}
\makeatother

\newcommand{\ep}{\epsilon}

\newcommand{\w}{\omega}
\newcommand{\E}{\mathbf{E}}
\newcommand{\B}{\mathbf{B}}
\newcommand{\J}{\mathbf{J}}

\begin{document}

\title{Hydrodynamic acoustic plasmon resonances in semiconductor nanowires and their dimers}

\author{Tahereh Golestanizadeh}
\affiliation{Department of Photonics Engineering, Technical University of Denmark, DK-2800 Kgs. Lyngby, Denmark}
\affiliation{Department of Physics, Yasouj University, Yasouj, 75914-353, Iran}
\author{Abbas Zarifi}
\email{Corresponding author: zarifi@yu.ac.ir}
\affiliation{Department of Physics, Yasouj University, Yasouj, 75914-353, Iran}
\author{Tahmineh Jalali}
\affiliation{Department of Physics, Persian Gulf University, Bushehr, 75196, Iran}
\author{Johan R. Maack}
\affiliation{Department of Photonics Engineering, Technical University of Denmark, DK-2800 Kgs. Lyngby, Denmark}
\author{Martijn Wubs}
\affiliation{Department of Photonics Engineering, Technical University of Denmark, DK-2800 Kgs. Lyngby, Denmark}
\affiliation{Center for Nanostructured Graphene (CNG), Technical University of Denmark}
\date{\today}

\begin{abstract}
The hydrodynamic Drude model known from metal plasmonics also applies to semiconductor structures of sizes in between single-particle quantum confinement and bulk. But contrary to metals, for semiconductors two or more types of plasma may have to be taken into account in order to properly describe their plasmonic properties.
In this combined analytical and computational study, we explore predictions of the recently proposed two-fluid hydrodynamic Drude model for the optical properties of plasmonic semiconductor nanowires, in particular for thermally excited InSb nanowires. We focus on the low-frequency acoustic surface and bulk plasmon resonances that are unique fingerprints for this model and are yet to be observed. We identify these resonances in spectra for single nanowires based on analytical calculations, and they are in complete agreement with our numerical implementation of the model. For dimers of nanowires we predict substantial increase of the extinction cross section and field enhancement of the acoustic localized surface plasmon resonance, which makes its observation in dimers more likely.
\end{abstract}

\maketitle

\section{Introduction}
Plasmonic structures have attracted quite some interest and found applications because of their ability to concentrate light on sub-wavelength scales and to greatly enhance electromagnetic fields~\cite{app5,app4,spp1,app3,sensing,app2}.
The Drude model is one of the most prevalent models to describe metal properties, but as the dimensions of metal structures approach the nanoscale, some plasmonic phenomena can no longer be explained by the classical Drude model.  
To explain these observed effects, such as the size-dependent blueshift of the resonance frequency of the localized surface plasmon (LSP) and confinement effects of bulk plasmons~\cite{Lindau,abajo1}, the semiclassical hydrodynamic Drude model (HDM) has often been used~\cite{HDM1,mcmahon,HDM4,HDM3,Intravia:2015a,Fitzgerald:2016a, HDM5,HDM6,Pitelet:2018a,Lelwala:2019a}.The linearized version of this model is similar to the Drude model, but with spatial dispersion included, which means that the polarization depends nonlocally on the electric field.

The HDM has recently been employed to describe other materials than metals that also have an electron plasma. 
Like for metals, it is well known that the plasmonic properties of bulk and large semiconductor  structures are well described using the Drude model~\cite{drude1,Hanham:2012a,Moridsadat:2018a,Kanyang:2018a,Sadrara:2019a}. In the other extreme case, semiconductor structures on the few-nanometer scale, such as quantum wells and dots, are described by quantum mechanical models for single- and few-electron excitations~\cite{quantum2,quantum}.
  For the intermediate regime, i.e. for sizes in between single-particle quantum confinement and bulk, the HDM model has recently been applied~\cite{johan2,deceglia,Johan}, and indeed, characteristic hydrodynamic resonances have been observed~\cite{deceglia} in semiconductors.
 Semiconductors have different bulk plasmonic properties than metals. It is well known that their lower charge carrier densities give rise to a lower plasma frequency, and therefore the frequencies of operation will be in the infrared or THz bands instead of the visible and near-UV regions as for metals. But because of the larger associated plasma wavelengths, nonlocal effects will generally be observable in larger structures for semiconductors than for metals. This can be convenient when dedicatedly fabricating structures in order to observe  their nonlocal effects~\cite{deceglia}, but it also means that semiconductor devices in general may exhibit unforeseen nonlocal behavior. 
 
The capability of adjusting and controlling charge carrier densities by varying temperature or doping levels are well known advantages of semiconductures~\cite{semi1},which allows for the design of tunable metamaterials see for exampleRefs.~\cite{Tanoto:2013a,Seren:2016a}. This tunability also facilitates the identifcation of nonlocal-reponse behavior in semiconductors, as we will see in this paper.
  The HDM considers only one of the possible kinds of charge carriers in semiconductors. Typically only electrons are included in the calculations, and this is a reasonable approximation when electrons are present as the majority carriers, since their smaller effective mass compared to the holes causes them to dominate the optical properties.

However, to account for the different kinds of charge carriers, an extension of the HDM that considers two different plasmas or fluids can be applied, and this is the idea of the recently proposed two-fluid hydrodynamic Drude model~\cite{Johan}.
In this model, two different branches of resonances are formed, optical and acoustic, both depending on properties of the electron and hole plasmas. Especially the acoustic branch is a unique signature of the two-fluid model~\cite{Johan} as it is absent in the single-fluid HDM. 
A different version of the two-fluid model has also been applied to electrolytes and ions~\cite{ion}, which are other examples of systems with multiple fluids.

In this paper, we will use the two-fluid model of Ref.~\cite{Johan} to study the optical properties of semiconductor monomers and dimers of nanowires. There are many actively studied and technologically important semiconductor nanowire materials (see for example the recent review by Joyce et al.~\cite{Joyce:2016a}), and we will focus on indium antimony (InSb) nanowires, which can be fabricated with high quality\cite{Plissard:2012a,Mourik:2012a}. We will initially consider the two-fluid model for monomers of nanowires, where we will see the appearance of acoustic plasmon resonances that are absent in the single-fluid HDM. The monomer geometry also allows us to compare our numerical implementation with the analytical solution, and we find an excellent agreement between the two. Then we present simulations of dimers of nanowires to study the consequences of the field enhancement in the gap between the nanowires, and we will show that the acoustic plasmonic resonance is enhanced for this dimer configuration compared to the monomer. This suggests a way for experimental confirmation of the two-fluid model using semiconductor nanowire dimers.  
With our hydrodynamic two-fluid model we are not the first to predict the existence of acoustic plasmons. Following the seminal work of Pines~\cite{Pines:1956a}, there have been many theoretical and experimental studies of acoustic plasmons, which have been reported to exist at metal surfaces and on 2D materials~\cite{Diaconescu:2007a,Park:2010a,Yan:2012a,Hrton:2012a,Politano:2018a}. For semiconductor structures, the situation is less clear. One of the advantages of our model compared to other more complex models is that it is simple enough to allow one to study complex plasmonic nanostructures where the acoustic plasmons may be enhanced.

\section{The two-fluid model and analytical results in k-space}
 The two-fluid hydrodynamic model considers two types of charge carriers or fluids, such as electrons and holes (or light and heavy holes) in a semiconductor. This model describes both plasmas, labelled $a$ and $b$, with a hydrodynamic equation of motion for the associated current densities $\J_{a,b}$, while the classical Maxwell wave equation describes the linear optical properties in these semiconductor structures. The governing equations for the two-fluid model are thus~\cite{Johan}:
 \begin{subequations}\label{coupled_equations}
\begin{equation}
\label{eq:fluid a}
\frac{\beta_a^2}{\omega^2+i\gamma_a\w}\nabla(\nabla\cdot\J_a)+\J_a=\frac{i\w\ep_0 \w_a^2 }{\omega^2+i\gamma_a\w}\E,
\end{equation}
\begin{equation}
\label{eq:fluid b}
\frac{\beta_b^2}{\omega^2+i\gamma_b\w}\nabla(\nabla\cdot\J_b)+\J_b=\frac{i\w\ep_0 \w_b^2 }{\omega^2+i\gamma_b\w}\E,
\end{equation}
\begin{equation}
\label{eq:maxwell}
-\nabla \times\nabla\times \E +\frac{\w^2}{c^2}\ep_{\infty}\E=-i\mu_0\w(\J_a+\J_b).
\end{equation}
\end{subequations}  
Here $\w_a$ and $\w_b$ are the plasma frequencies, $\gamma_a$ and $\gamma_b$ are the damping constants, and $\beta_a$ and $\beta_b$ are the nonlocal parameters for the fluids $a$ and $b$, respectively. From Eq.~(\ref{coupled_equations}) it is clear that both plasmas are driven by the same electric field, while in turn the electric field is driven by the sum of the two current densities.
To obtain our analytical results for nanowires below, we will make use of the longitudinal and transverse wavenumbers in the two-fluid model~\cite{Johan} that we briefly present here. By transforming Eq.~(\ref{coupled_equations}) to k-space, two different longitudinal wavenumbers (both denoted by the subscript $'L'$) are obtained with dispersion relations 
\begin{subequations}\label{eq:long_wavenumber}

\begin{equation}
\label{eq:K}
k^2_{L,j} =\frac{1}{2}\left(k_a^2+k_b^2 \pm \sqrt{(k_a^2-k_b^2)^2+\frac{4\w_a^2\w_b^2}{\beta_a^2\beta_b^2\ep^2_{\infty}}}\  \right) j=1,2,
\end{equation}

where
\begin{equation}
\label{eq:k_i}
k_i^2 =\left(\w(\w+i\gamma_i)-\frac{\w_i^2}{\ep_{\infty}}\right)\frac{1}{\beta_i^2}\qquad i=a,b.
\end{equation}
\end{subequations}
These two longitudinal modes are the result of hybridization of the two different kinds of charge carriers, and the dispersion relations are displayed in Fig.~1 of Ref.~\cite{Johan}.
Here we review the properties of the two modes in some detail. We define $k_{L,2}(\w)$ to be the minus ('$-$') solution in Eq.~(\ref{eq:K}), and it corresponds to an optical longitudinal mode, analogous to the optical longitudinal mode at the plasma frequency within the local Drude model. The single-fluid hydrodynamic model also has one such an optical longitudinal mode. The lowest frequency at which the optical longitudinal mode exists within the two-fluid model is $\omega_{\rm eff}/\sqrt{\epsilon_{\infty}}$, where we have introduced the effective bare plasma frequency $\omega_{\rm eff} = ({\omega_{a}^{2} + \omega_{b}^{2})^{1/2}}$. This minimal frequency can be found by putting $k_{L,2}(\w)$ to zero while neglecting $\gamma_{i}$ in Eq. (\ref{eq:K}). Below this frequency, $k_{L,2}(\w)$ becomes purely imaginary which means that no propagating longitudinal optical modes exist.
But as mentioned in the Introduction, the decisive new feature for the two-fluid model as compared to both the single-fluid HDM and the local Drude model is the existence of an acoustic longitudinal mode. Its wavevector is denoted by $k_{L,1}(\w)$ which corresponds to the '$+$' solution of Eq.~(\ref{eq:K}). In stark contrast to the optical mode, this acoustic mode exists all the way down to zero frequency, as can be found by putting $k_{L,1}(\w)$ to zero. Therefore the two-fluid model exhibits longitudinal acoustic excitations and resonances in the low-frequency region below $\omega_{\rm eff}/\sqrt{\epsilon_{\infty}}$ where no longitudinal optical modes exist.  
 The additional low-frequency resonances are a characteristic feature of the two-fluid model. 

Besides the longitudinal wavenumbers, there is also a the transverse wavenumber given by 
\begin{equation}
\label{eq:epsilonT}
k_{T}^2 = \frac{\w^2}{c^2} \left(\ep_{\infty}-\frac{\w_a^2 }{\w^2+i\gamma_a\w}-\frac{\w_b^2 }{\w^2+i\gamma_b\w}\right),
\end{equation}
featuring two Drude-like response terms. The plasma frequencies and damping constants, which also were used in Eqs.~(\ref{coupled_equations}) and (\ref{eq:long_wavenumber}), are given by \cite{cond}
\begin{equation}
\label{eq:omega, gammaa}
\w_i^2=\frac{e^2 n_i}{\ep_0 m^*_i} ,
\qquad
\gamma_i=\frac{e}{m^*_{i,{\rm cond}} \mu_i}.
\end{equation}
Here $n_i$ and $m^*_i$ are the charge-carrier densities and the density-of-states effective masses, respectively, and the $m^*_{i,{\rm cond}}$ are the conductivity effective masses. The damping constants $\gamma_i$ depend furthermore on the respective charge-carrier mobilities $\mu_{i}$.
The nonlocal parameters $\beta_i$ depend on the nature of the charge carriers, and for thermally excited charge carriers in an intrinsic semiconductor they are given by~\cite{Johan,johan2}
\begin{equation}
\label{eq:beta}
\beta_i^2=\frac{3k_B T}{m^*_i},
\end{equation}
where $T$ is the temperature, and $k_B$  the Boltzmann constant. This expression is only valid for temperatures low enough that the Fermi-Dirac distribution can be approximated by the Boltzmann distribution.
\subsection{Two-fluid model for a single nanowire: analytical results}\label{Sec:wireanalytical}
In this paper, we will analyze the two-fluid model for finite systems by focusing on semiconductor nanowires and, to begin with, a single wire. Let us first mention that, recently, Mie-coefficients for the two-fluid model in spherical geometry were derived~\cite{Johan}, where in comparison to HDM a second longitudinal wave was accounted for, namely the acoustic one. Analytical solutions exist for infinite cylindrical wire geometries, and for the Drude model, these can be found in textbooks~\cite{bohren}. For the linearized (single-fluid) HDM, analytical solutions were obtained both in the quasi-static limit~\cite{VilloPerez:09} and in the fully retarded case~\cite{ruppin,dtu5,dtu3}, and recently also for the full nonlinear HDM~\cite{Moeferdt:2018a}. For the linearized two-fluid model, analytical solutions have not yet been obtained for nanowires, and we will therefore derive them here.  

In particular, we will calculate the extinction cross section for a single nanowire of radius $R$, by assuming an incident plane wave with the electric field polarized perpendicular to the cylinder axis (TM modes), since longitudinal modes may only be excited by this polarization (and not by TE modes). The incident field $\E_i$ and the resulting scattered (reflected) wave from the nanowire $\E_r$ are given by~\cite{bohren}
\mathleft
\begin{subequations}
\begin{align}
\label{eq:E_i}
&\E_i\!=\!\frac{i E_0}{k_D}\!\!\!\sum_{n=-\infty}^\infty\!\!\!\! i^n \!\!\left(\!\frac{in}{r}J_n(\! x_D\!)e^{in\theta} \hat{\mathbf{r}}\!-\! k_DJ_n^\prime(\! x_D\!)e^{in\theta}\hat{\bm{\theta}}\!\right)\! , \\
\label{eq:E_r}
&\E_r\!=\!\frac{i E_0}{k_D}\!\!\!\sum_{n=-\infty}^\infty\!\!\!\! a_n i^n \!\!\left(\!\frac{in}{r}H_n^{(1)}(\! x_D\!)e^{in\theta}\hat{\mathbf{r}}\!-\! k_D H_n^{(1)\prime}(\! x_D\!)e^{in\theta}\hat{\bm{\theta}}\!\right)\!,
\end{align}
\end{subequations}
\mathcenter
where the $J_n$ and $H_n^{(1)}$ are the Bessel and Hankel functions of the first kind, respectively. Here $x_D  \equiv R k_D$ and $k_D \equiv \sqrt{ \ep_D} \w/c$, where $\ep_D$ is the permittivity of the surrounding dielectric. The coefficients $ a_n $ of the scattered wave are still to be determined.
The transmitted wave into the nanowire (inside the cylinder) $\E_t$ contains, in addition to the transversal field, two different longitudinal fields, such that
$\E_t= \E^{\rm T}+\E^{\rm L}$ with 
\mathleft
\begin{subequations}\label{eq:ETandELinside}
\begin{align}
  \label{eq:E_T}
 \E^{\rm T}\!=&\frac{i E_0}{k_T}\!\!\!\sum_{n=-\infty}^\infty\!\!\!\! g_n i^n \!\!\left(\!\frac{in}{r}J_n(\! x_D\!)e^{in\theta} \hat{\mathbf{r}}\!-\! k_T J_n^\prime(\! x_D\!)e^{in\theta}\hat{\bm{\theta}}\!\right)\!,\!\!\! \\
\label{eq:E_L}
\E^{\rm L}\!=&i E_0 \!\!\!\sum_{n=-\infty}^\infty\!\!\!\! \left(\! \frac{1}{k_{L,1}}h_{1n} i^n \!\!\left(\!k_{L,1} J_n^\prime(\! x_1\!)e^{in\theta}\hat{\mathbf{r}}\!+\!\frac{in}{r}J_n(\! x_1\!)e^{in\theta} \hat{\bm{\theta}}\!\right)\right.\nonumber \\ 
 +&\left.\frac{1}{k_{L,2}} h_{2n} i^n \!\!\left(\!k_{L,2} J_n^\prime(\! x_2\!)e^{in\theta}\hat{\mathbf{r}}\!+\!\frac{in}{r}J_n(\! x_2\!)e^{in\theta} \hat{\bm{\theta}}\!\right)\!\right)\!,\!\!\!
\end{align}
\end{subequations}
\mathcenter
with expansion coefficients $g_n$ for the transverse waves and $ h_{1n}$ and $h_{2n}$ for the two longitudinal waves that are also still to be determined. 
The differentiation (denoted with the prime) is with respect to the argument. 
The definitions $x_T  \equiv Rk_T$ and $x_j  \equiv R k_{L,j}$ have been used in Eq. (\ref{eq:ETandELinside}), where $k_T$ and $k_{L,j}$ are given by Eq.~(\ref{eq:epsilonT}) and Eq.~(\ref{eq:K}), respectively. So in the cylinder the same optical and acoustic longitudinal waves exist as in the infinite two-fluid medium, and their amplitudes $h_{1n}$ and $h_{2n}$ quantify how well these waves can be externally excited by a plane wave and resonate within the confines of the cylinder. 
The unknown wave amplitudes are determined by applying boundary conditions. 
The continuities of $\E_ \parallel $ and $ \B_ \parallel$ across boundaries are consequences of Maxwell's equations and provide us with the first two boundary conditions.The 'hard-wall' additional boundary condition (ABC) $ \J_ \perp =0 $, which implies that the charge carriers cannot leave the surface, is used widely for the HDM. The two-fluid model requires two ABCs, and here we will follow Ref.~\cite{Johan} and use the conditions $\J_{a,\perp } =\mathbf{0}$ and $\J_{b,\perp} =\mathbf{0}$.
   We hereby obtain the linear system of equations as presented in Appendix~A, from which all coefficients can be found. Of primary interest in order to calculate the extinction cross section are the coefficients $a_n$, which are given by 
\mathleft
\begin{equation}
\label{eq:a}
a_n\!\!=\!\!\frac{-\sqrt \ep_D J_n(\! x_D\!)[-J_n^\prime(\! x_T\!)\!+\! \Delta_n]\!-\!\sqrt \ep_T J_n(\! x_T\!)J_n^\prime(\! x_D\!)}{\sqrt \ep_D H_n^{(1)} \!(\! x_D\!)[-J_n^\prime(\! x_T\!)\!+\! \Delta_n]\!+\!\sqrt \ep_T J_n(\! x_T\!) H_{n}^{(1)\prime}\!(\! x_D\!)}.
\end{equation}
\mathcenter
Here the parameter $ \Delta_n$ accounts for the nonlocality of both plasmas and is given by
\begin{equation}
\label{eq:delta}
\Delta_n= \frac{J_n(x_T)n^2}{x_T A}\left(\frac{J_n(x_1)C_2}{x_1J_n^\prime (x_1)}-\frac{J_n(x_2)C_1}{x_2J_n^\prime (x_2)}\right),
\end{equation}
where $A$ and $C_j$ are given by
\begin{subequations}\label{A,C, alpha}
\begin{equation}
\label{eq:A}
A=\frac{\ep_{\infty}^2 (\omega^2+i\gamma_a\w)(\omega^2+i\gamma_b\w)}{\beta_a^2\beta_b^2(1+\alpha_1)(1+\alpha_2)}(\alpha_2^L-\alpha_1^L),
\end{equation}
\begin{equation}
\label{eq:C}
C_j= \frac{\w_a^2\ep_{\infty}k_{L,j}^2 }{\beta_a^2 (1+\frac{1}{\alpha_j^L})} -\frac{\w_b^2\ep_{\infty}k_{L,j}^2 }{\beta_b^2 (1+\alpha_j^L)},
\end{equation}
\begin{equation}
\label{eq:alpha}
    \alpha_j^L= \frac{\beta_a^2\ep_{\infty} }{\w_a^2}(k_a^2-k_{L,j}^2).
\end{equation}

\end{subequations}
In the limit $\Delta_n \to 0$, the cylinder scattering coefficient $a_n$ indeed reduces to the local-response expression. We are interested in the extinction cross section per unit length of the cylinder $\sigma_{\rm ext}$,which becomes the dimensionless normalized cross section $C_{\rm ext}$ after dividing by the wire diameter. In terms of the amplitudes $a_n$, this normalized extinction cross section is given by~\cite{ruppin}
\begin{equation}
\label{eq:sigma}
C_{\rm ext} = -\frac{2}{k_DR}\sum_{n=-\infty}^\infty\mathrm{Re}(a_n).
\end{equation}
This expression is valid for all the models that we consider; the differences between the models show up as different expressions for the coefficients $a_n$. 
\section{Two-fluid model for a single nanowire: numerical results and benchmark}\label{sec:???}
To investigate the properties of the two-fluid model, we start with analyzing an artificial semiconductor material where damping constants have been set low to easily identify the individual characteristic resonances. After that we will consider the intrinsic semiconductor indium antimonide (InSb). 

We calculate the spectrum for a single nanowire both analytically [using Eq.~(\ref{eq:sigma})] and numerically, which will also benchmark our numerical implementation. Numerical calculations are performed using a commercially available finite-element-based package (COMSOL~5.3a). Our implementation of the two-fluid model is a generalization of the one for the HDM by Toscano et al.~\cite{dtu3}, for which the code is freely available~\cite{nanopl:2012a}. In our simulations in two dimensions, the nanowire is embedded into a big rectangular computational domain of air, and the structure is terminated by perfectly matched layers (PMLs) to provide an approximately reflection-free termination of the free-space domain. The extinction cross section is obtained by numerically integrating the Poynting vector on a circle surrounding the nanowire. 
\begin{figure}
 \centering\includegraphics[width=.4\textwidth]{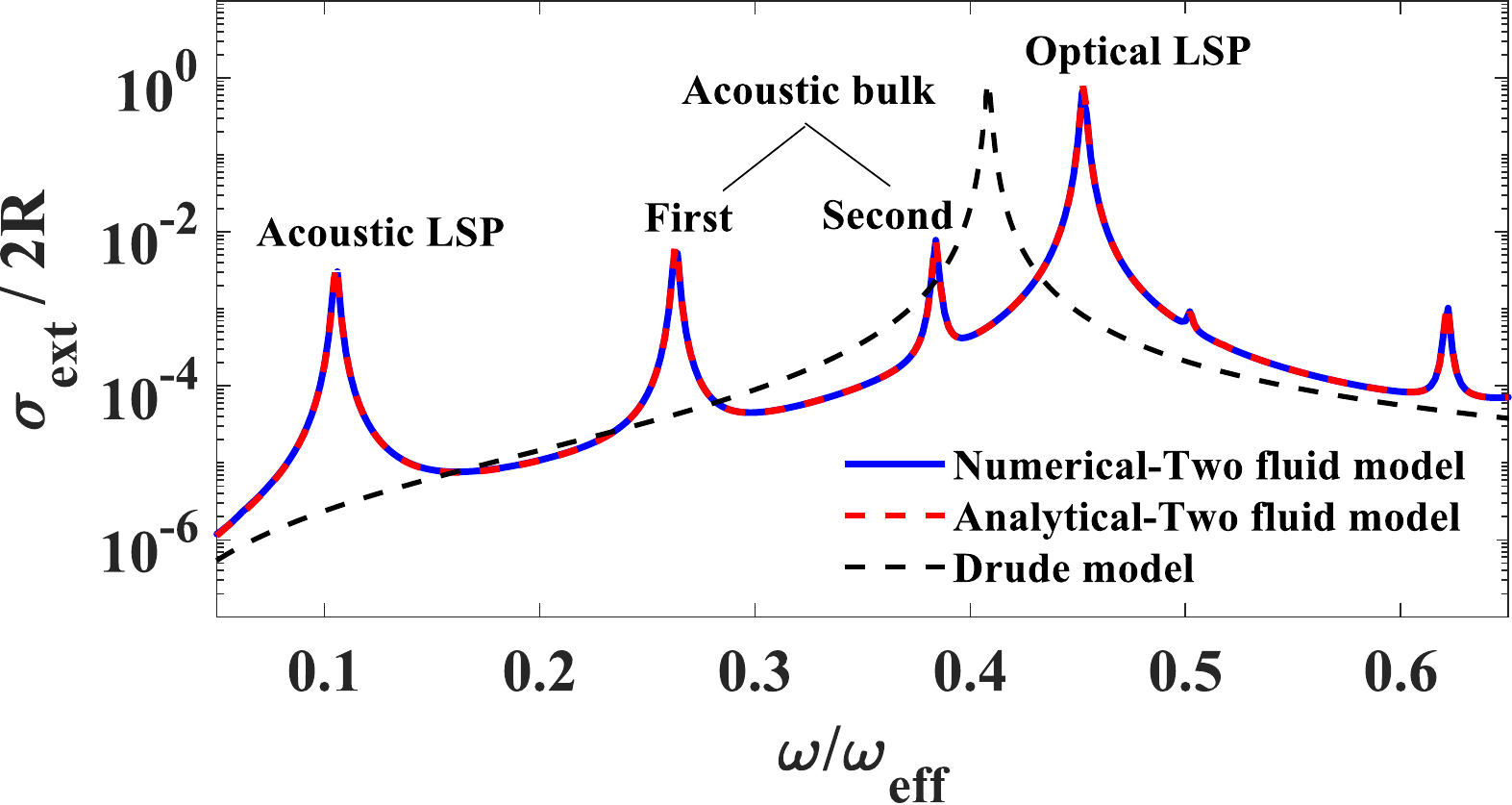} 
\caption{The extinction spectrum of a semiconductor nanowire with parameters $\omega_a=3.6\times10^{14} \mathrm{\,s^{-1}}$, $\omega_b=1.8\times10^{14}\mathrm{\,s^{-1}} $, $\gamma_a=\gamma_b=1.0\times10^{12}\mathrm{\,s^{-1}}$,
 $\beta_a=4.3\times10^{5}\mathrm{\,m\,s^{-1}}$,
  $\beta_b=1.6\times10^{5}\mathrm{\,m\,s^{-1}}$ and $\epsilon_\infty=5$, with $R=10\mathrm{\,nm}$ and $\epsilon_D=1$. The solid, blue line is the numerical solution of the two-fluid model, and it overlaps the corresponding analytical solution shown with the dashed, red line. For comparison, the local Drude model is shown with the dashed, black line. In all cases, the incident light is a TM-polarized plane wave normally incident on the nanowire.}
\end{figure}

 \begin{figure}
 \centering\includegraphics[width=.4\textwidth]{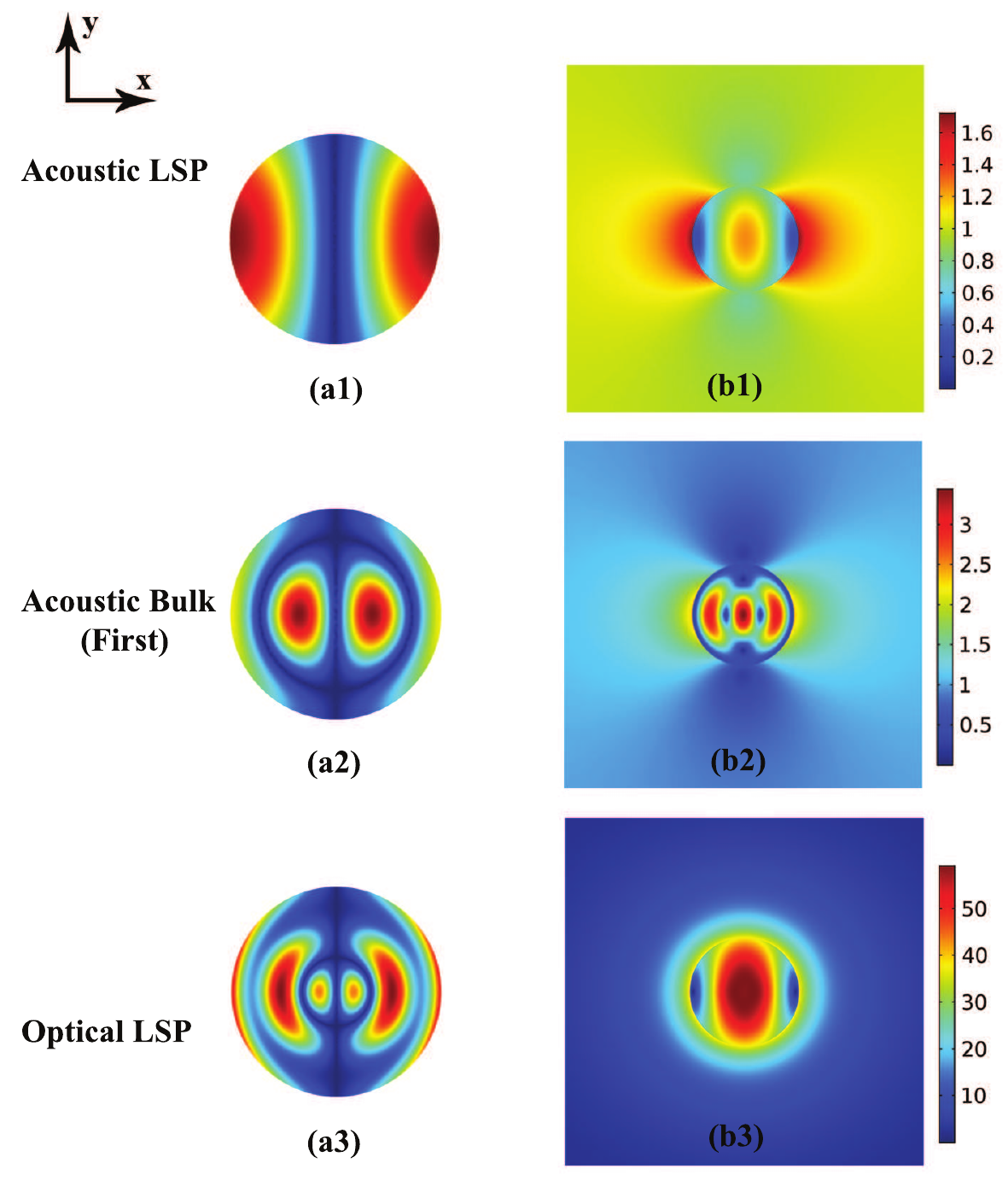}
\caption{Absolute values of the charge distribution and the electric-field distribution of the same nanowire as in Fig.~1. The left panels labeled~(a) depict the charge distributions ($\mathrm{C/m^3}$), and panels on the right labeled~(b) show the norm of the electric field ($\mathrm{V/m}$) for three mentioned modes. The incident electric field is polarized in the $x$-direction.}
\end{figure}

Figure~1 shows spectra for a single nanowire with radius $R=10~\mathrm{\,nm}$ and with further parameters given in the caption. The spectrum in Figure~1 has been normalized to the diameter of the nanowire and is presented as a function of the scaled frequency $\omega/\omega_\mathrm{eff}$. 
Clearly, in Fig.~1 the numerical and analytical spectra for the two-fluid model overlap completely. The nanowire thus serves as a benchmark problem that shows that our numerical implementation of the model is reliable. This is, as far as we know, the first numerical implementation of the two-fluid model that was introduced in Ref.~\cite{Johan}, and it will allow us in Sec.~\ref{Sec:wiredimers} to employ the numerical implementation with confidence, also for geometries more complex than single wires. 
The black dashed line in Fig.~1  is the local-response approximation obtained by setting $\Delta_n$ equal to zero in Eq.~(\ref{eq:a}). This spectrum only shows a single peak, namely the dipole LSP peak.   
It is well known for the single-fluid HDM that the LSP resonance blueshifts compared to the local-response model~\cite{scholl}, and recently it was found that the relative nonlocal blueshift for semiconductors typically is much larger than for metals~\cite{johan2}. Here we see again a large blueshift of the optical LSP peak (compare the highest peaks in the Drude and two-fluid spectra) for nanowires in the two-fluid model. We checked that the blueshift is larger for smaller wire radii (not shown in Fig.~1), which is indeed the expected behavior in this hard-wall hydrodynamic model.  
Neither in the LRA nor in the single-fluid HDM are there plasmonic resonances of a single plasmonic particle below the dipole LSP peak. 
But for the two-fluid model, several such peaks are visible in the spectrum of Fig.~1, and they are thus characteristic of the two-fluid model. From our discussion of the modes it is clear what these resonances are: they are resonances of the acoustic longitudinal mode that for the two-fluid model exists down to zero frequency. For this reason they can be called acoustic peaks, as is done in Fig.~1. Analogous resonances of acoustic modes in singlespheres were reported in Ref.~\cite{Johan}.
For selected resonant modes in the spectrum of Fig.~1, we depict in Fig.~2 the absolute values of the charge distribution and the electric-field distribution of the nanowire in the left and right panels, respectively. As before, the incident light is a TM-polarized plane wave normally incident on the wire, with the electric field polarized in the $x$-direction.

 Figure~2(a1) corresponds to the first peak of the spectrum in Fig.~1, and we see a high charge density near the surface, which indicates that this is a LSP mode. So we identify and henceforth call this first peak 'acoustic LSP'. Fig.~2(a2) is the charge distribution of the second peak, and we see that this is a bulk plasmon with a high charge density near the center. 
 The charge distribution for the dipole LSP peak is shown in Fig.~2(a3). We will henceforth call this peak 'optical LSP', and although it is a surface plasmon, we see from the figure that is also has the signature of a bulk plasmon. This is caused by a mixing of the LSP and the bulk plasmon nearby.
  In Fig.~(2b) we depict the electric-field distributions for the discussed modes, and in Fig.~2(b3) one can see field enhancement both at the wire edge and in its center, illustrating the hybrid character with bulk and localized surface plasmon characteristics combined. 
  After studying the different aspects of the two-fluid model for a nanowire with artificially low damping, we will now consider a realistic semiconductor and choose intrinsic InSb with thermally excited charge carriers.
\begin{table}
\caption{Material parameters for InSb at various temperatures. The sources of the values can be found in the rightmost column. Note that the conductivity effective mass for the electrons $m_{e,\mathrm{cond}}^*$ is assumed to be equal to the density-of-states effective mass $m_e^*$, and that the value for $m_{lh}^*$ is for $T = 20\mathrm{\,K}$~\cite{madelung}.}
\scalebox{0.8}{%
 \begin{tabular}{lrrrrrc}
 \hline
  & $T=200\mathrm{\,K}$ & $T=300\mathrm{\,K}$ & $T=350\mathrm{\,K}$ & $T=400\mathrm{\,K}$ & Refs. \\
 \hline
$\epsilon_\infty$ & $15.68$ & $15.68$ & $15.68$ & $15.68$ & \cite{madelung} \\
$E_g$ [$\mathrm{eV}$] & $0.200$ & $0.174$ & $0.160$ & $0.146$ & \cite{rowell88} \\
$\mu_e$ [$\mathrm{cm^2/V s}$] & $151000$ & $77000$ & $60000$ & $48000$ & \cite{madelung} \\
$\mu_h$ [$\mathrm{cm^2/V s}$] & $1910$ & $850$ & $620$ & $480$ & \cite{madelung} \\
$m_e^*/m_0$  & $0.0125$ & $0.0115$ & $0.0108$ & $0.0100$ & \cite{stradling70} \\
$m_{e,\mathrm{cond}}^*/m_0$  & $0.0125$ & $0.0115$ & $0.0108$ & $0.0100$ & \cite{stradling70} \\
$m_{lh}^*/m_0$  & $0.016$ & $0.016$ & $0.016$ & $0.016$ & 
\cite{madelung} \\
$m_{hh}^*/m_0$  & $0.37$ & $0.37$ & $0.38$ & $0.40$ & \cite{oszwaldowski87} \\ 
\hline
\end{tabular}}
\end{table}
\begin{figure}
\centering\includegraphics[height=8cm,width=.4\textwidth]{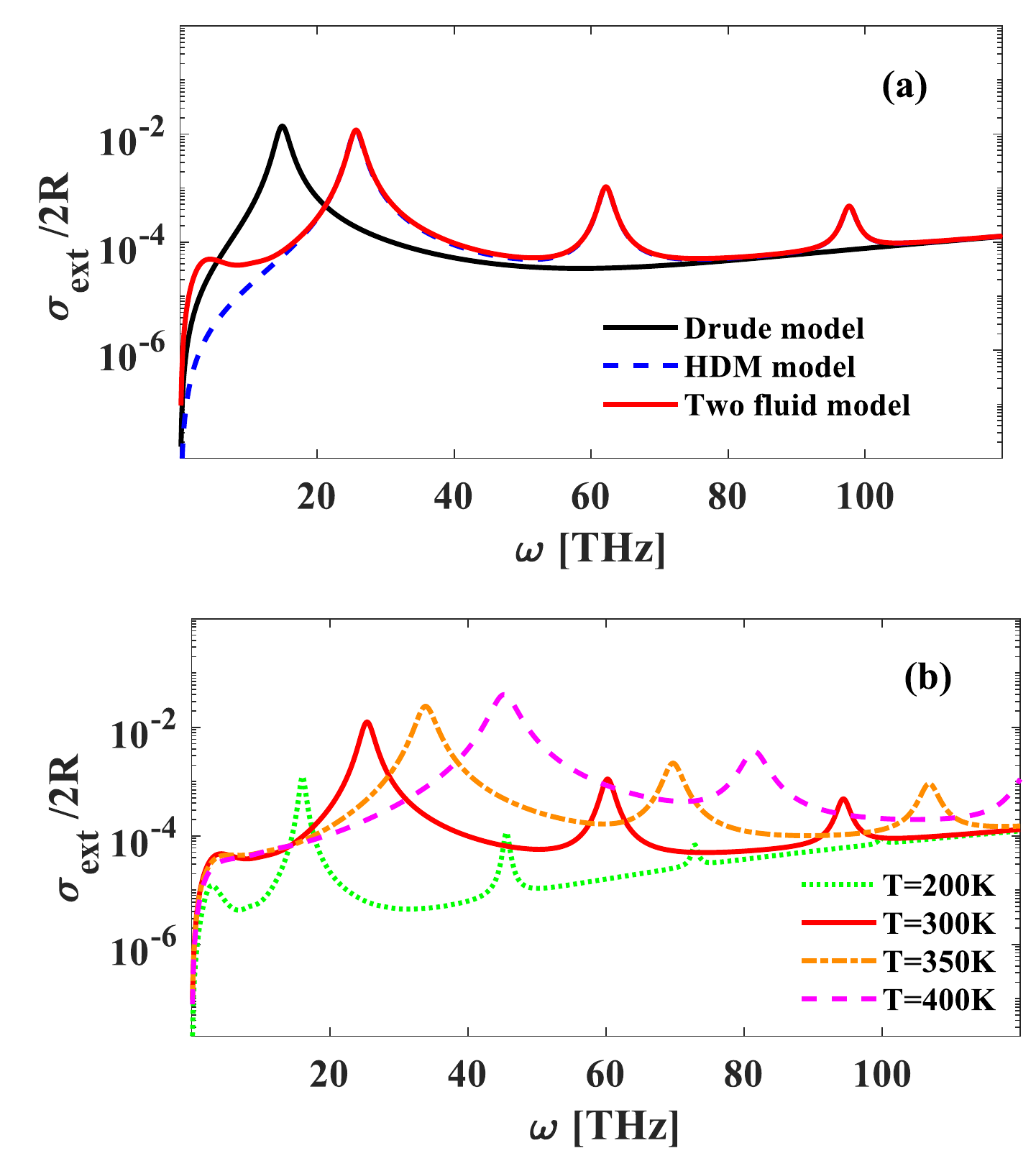}
\caption{ Extinction spectra for a single nanowire in vacuum with $R=100\mathrm{\,nm}$ made of intrinsic InSb. See Table~1 and main text for the material parameters of the InSb. (a) At $T=300\mathrm{\,K}$ for three models: the black line is the local Drude model, the red line is the two-fluid model and the blue dashed line is the (single-fluid) HDM. 
(b) Comparison of extinction spectra at $T=200\mathrm{\,K}$, $300\mathrm{\,K}$, $350\mathrm{\,K}$ and $400\mathrm{\,K}$. } \end{figure} 

For the InSb nanowire we take $R=100\;\mathrm{\,nm}$ and  $T=300\mathrm{\,K}$, with electrons ($e$) as the $a$-fluid and holes ($h$) as the $b$-fluid. We then use Eqs.~(4) and (5) with data from Table~1. This gives us $\omega_e=6.328\times10^{13} \mathrm{\,s^{-1}} $, $\omega_h=1.115\times10^{13}\mathrm{\,s^{-1}}$, $\gamma_e=1.990\times10^{12}\mathrm{\,s^{-1}}, \gamma_h=6.674\times10^{12}\mathrm{\,s^{-1}}$,
 $\beta_e=1.090\times10^{6}\mathrm{\,m\,s^{-1}}$ and $\beta_h=1.920\times10^{5}\mathrm{\,m\,s^{-1}}$. 
In Fig.~3(a) we compare the corresponding extinction spectra for three models: The Drude model, the HDM and the two-fluid model. The spectrum for the single-fluid HDM is obtained by including only electrons in the calculation, which means that the parameters are given by $\omega_p=\omega_e$, $\beta=\beta_e$ and $ \gamma=\gamma_e$. As we expect, the Drude model only results in one peak: the dipole LSP, just like in Fig.~1. The blueshift of the corresponding resonance in both nonlocal models is substantial and practically identical in the two cases.  

The optical LSP peak is followed by several bulk plasmon peaks, the first two of which are shown in Fig.~3(a). Comparison to Fig.~1 for the artificial semiconductor reveals that only one of the acoustic peaks is visible at low frequency, and the others are practically absent mainly due to the low mobility of the holes. In particular, only the low-frequency acoustic LSP peak is visible in Fig.~3(a) while the acoustic bulk peaks are not. So the noteworthy difference between the two nonlocal models (two-fluid model and HDM) for the InSb nanowire is the existence of a LSP acoustic peak in the two-fluid model below the optical LSP peak.
At higher frequencies, the holes and electrons de-hybridize~\cite{Johan} and the extinction curves of the two nonlocal models overlap almost completely.
In Fig.~3(b) we compare the extinction spectra for various temperatures ( $T=200\mathrm{\,K}$, $300\mathrm{\,K}$, $350\mathrm{\,K}$ and $400\mathrm{\,K}$) for the same nanowire as in Fig.~3(a). At higher temperatures the peaks broaden and the larger free-carrier densities shift all peaks to the blue, and we find to an overall increase of the extinction cross section. 
\section{ Dimers of cylindrical nanowires}\label{Sec:wiredimers}
 Here we study within the two-fluid model both the extinction cross section and the electric-field intensity enhancement of dimers of cylindrical nanowires. This study is enabled by our numerical implementation of the model that we validated in the previous section. We consider a dimer of the same artificial low-loss material as for the nanowire in Fig.~1. As an archetypical plasmonic dimer structure, we consider two identical and parallel cylindrical nanowires, with the same radii of $R=10\mathrm{\,nm}$ and separated by a gap distance of $d=1\mathrm{\,nm}$. Extinction spectra normalized to the diameter of the nanowire are shown in Fig.~4a, where the different excitation directions, along and normal to dimer axis, are compared to a monomer identical to the single nanowire of Fig.~1.
\begin{figure}
\centering\includegraphics[height=8cm,width=.4\textwidth]{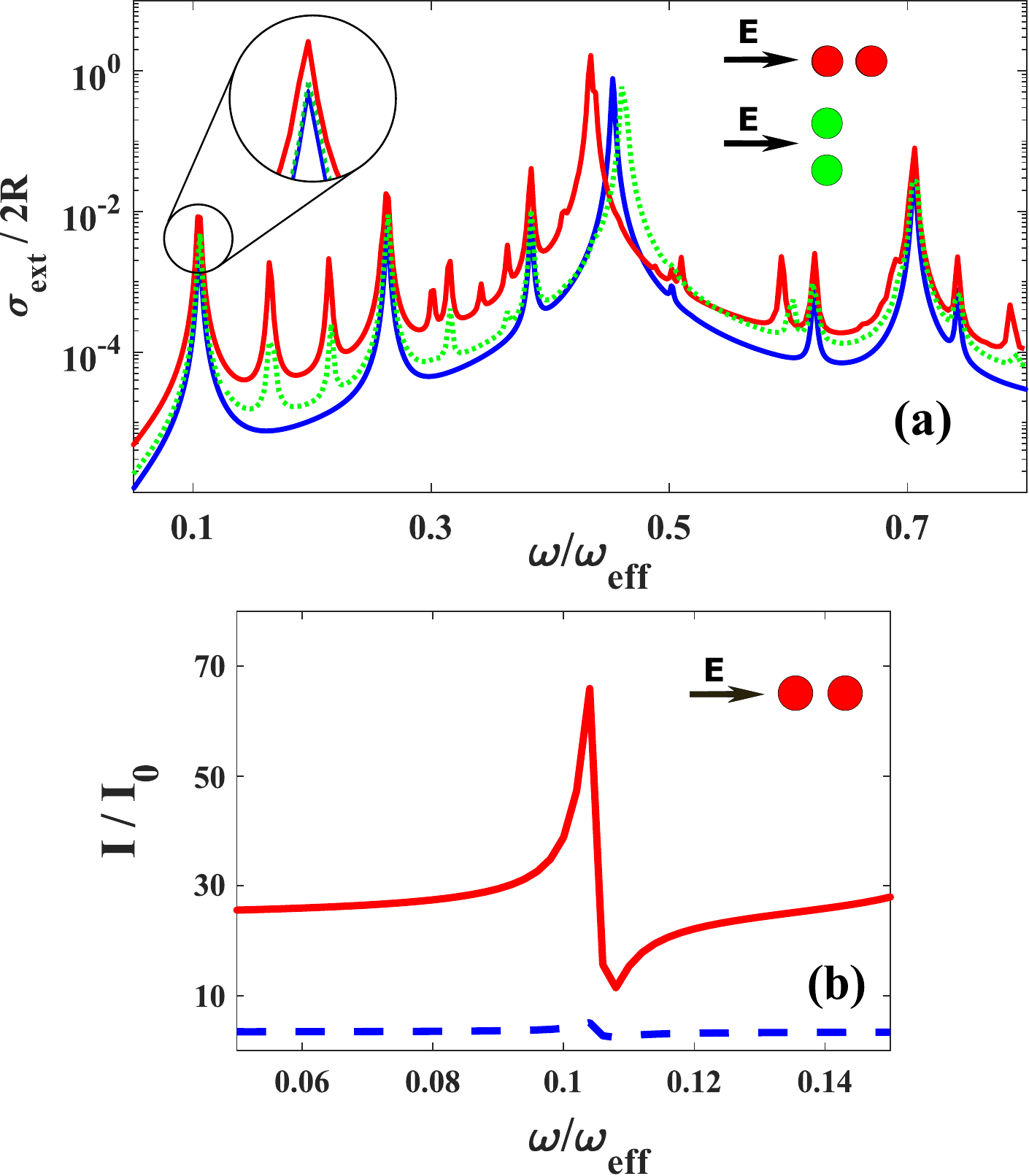}
\caption{(a) Extinction spectra for a semiconductor dimer of $R=10\mathrm{\,nm}$ and $d=1\mathrm{\,nm}$ (same material as in Fig.~1). The blue curve is for the monomer, while the green and red curves correspond to dimers excited by a normally incident plane wave with the electric field normal to and along the dimer axis, respectively. (b) Normalized intensities of monomer and dimer in the vicinity of the acoustic LSP. The intensity for the dimer is evaluated in the middle of the gap; for the monomer the value is taken at the same position with the right cylinder removed.}
\end{figure} 
 Fig.~4(a) shows that the incident wave with electric field polarized parallel to the dimer axis gives a stronger acoustic resonance than for the perpendicular polarization direction. 
In that sense the acoustic modes behave similar to the well-known optical LSP modes in dimers, but the enhancement of the acoustic mode is not as pronounced (comparison not shown).
Another aspect worth mentioning is that the spectral position of the acoustic LSP mode in Fig.~4(a) is practically constant in all studied cases (monomer, and dimer excited in the two directions), whereas the position of the optical LSP mode of the dimer (highest peak of green and red curve) is considerably shifted with respect to the (blue) monomer resonance. One could interpret this as a complete absence of hybridization of the acoustic LSP modes of the two cylinders, but this conclusion is too rash, as Fig.~4(b) will illustrate.

Fig.~4(b) shows the corresponding intensity normalized to the incident intensity, close to the acoustic LSP peak. The excitation is polarized along the dimer axis,  as it was for the red curve in Fig.~4(a). We see how the dimer increases the electric-field intensity in the narrow gap between the two nanocylinders. The increase is by a factor of 13 in comparison to a single nanowire at same distance. 
This shows that in the experimental pursuit of acoustic plasmon resonances, it can be advantageous to study dimers, where these resonances are stronger. It also suggests that the acoustic LSP modes of the two cylinders do hybridize, but not strongly enough so as to produce an observable resonance shift in Fig.~4(a).  
  \begin{figure}
\includegraphics[height=8cm,width=.4\textwidth]{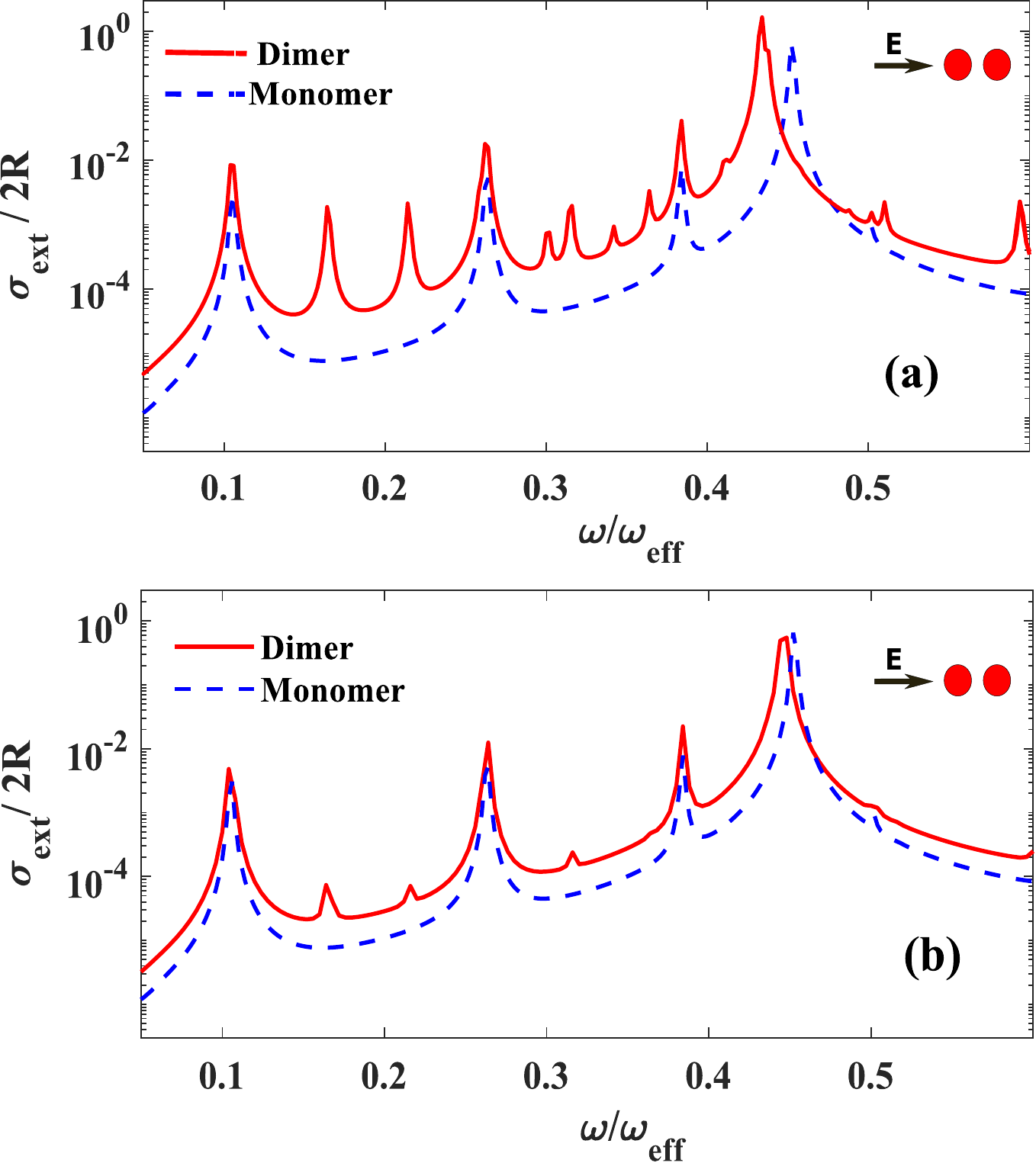}
\caption{Extinction spectra for a semiconductor dimer of $R=10\mathrm{\,nm}$ with a gap size of (a) $d=1\mathrm{\,nm}$ and (b) $d=10\mathrm{\,nm}$ compared with a  monomer. The material parameters are the same as for the monomer in Fig.~1.}
\end{figure} 

In Fig.~5 we illustrate the effect of the gap distance for the same type of dimers as above. 
Comparison of panels~(a) and (b) shows the familiar increasing hybridization redshift of the optical LSP as the gap distance is reduced~\cite{redshift}. By contrast, the spectral position of the acoustic peak appears to be independent of the gap distance. The latter is consistent with our findings in Fig.~4: if a gap distance of $1\mathrm{\,nm}$ (in Fig.~4 and 5(a)) is not small enough to give a hybridization of the acoustic LSPs leading to an observable peak shift compared to the monomer, then the gap distance of $10\mathrm{\,nm}$ in Fig.~5(b) will surely not show such a shift. The enhancement of the acoustic resonance, however, for the dimer as compared to the monomer is larger in Fig.~5(a) than in~5(b), as expected.

 Finally we study the influence of gap size and of temperature on dimer spectra for InSb, as these are variables that may be tuned in experiments. We keep the radius fixed at $R=100\mathrm{\,nm}$ and consider an incoming wave with the electric field polarized parallel to the dimer axis. 
 In Fig.~6(a) the extinction spectra normalized to the diameter of the nanowire are shown for different gap sizes at a constant temperature of $T=300\mathrm{\,K}$, while in  Fig.~6(b) we compare the extinction spectra for various temperatures while keeping the gap distance at $d=10\mathrm{\,nm}$.

\begin{figure}
\centering
\includegraphics[height=8cm,width=.4\textwidth]{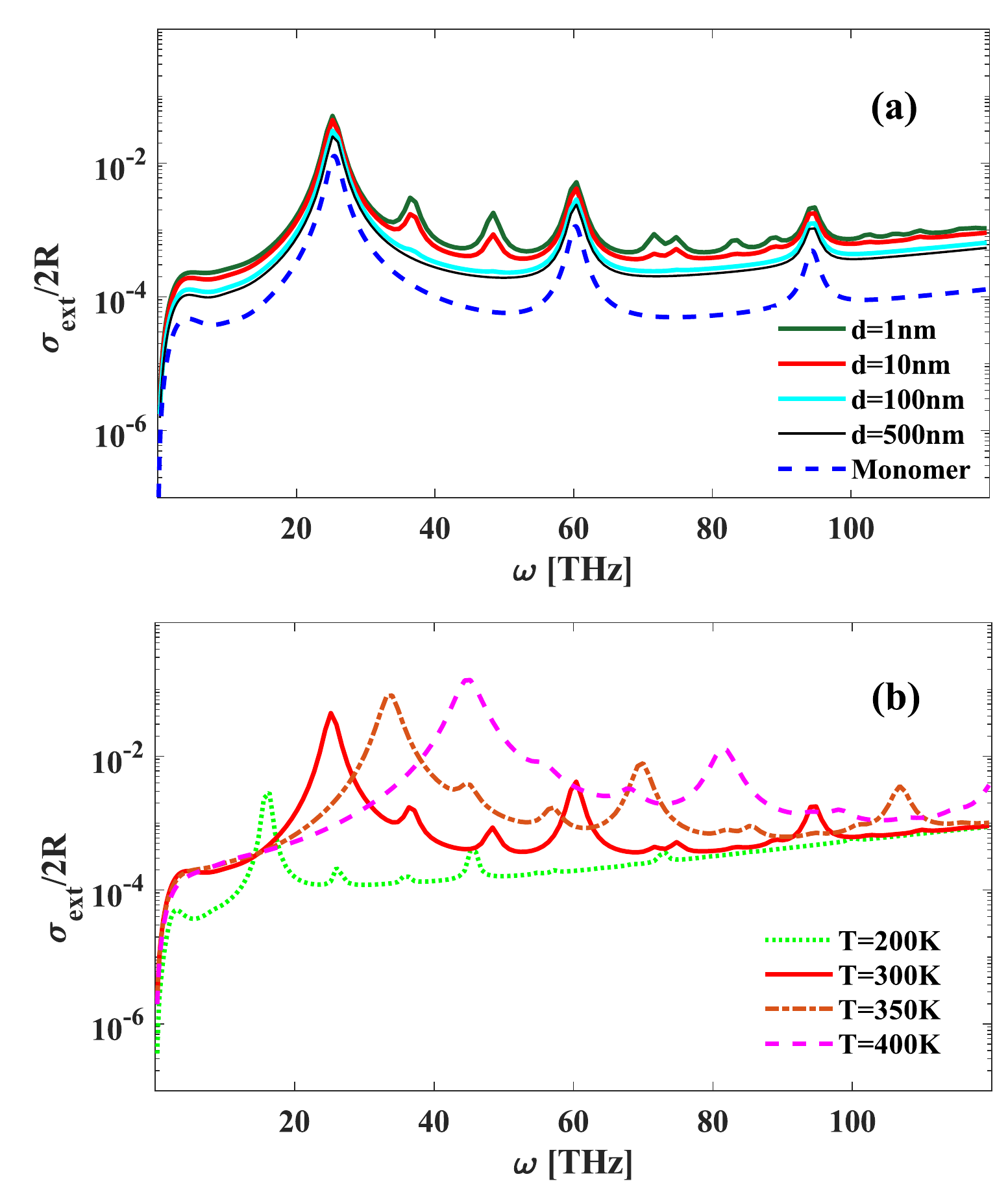}

\caption{(a) Extinction spectra for nanowire dimers of intrinsic InSb with radius $R=100\mathrm{\,nm}$ in vacuum with gap sizes $d=1\mathrm{\,nm}$, $10\mathrm{\,nm}$, $100\mathrm{\,nm}$ and $500\mathrm{\,nm}$ at $T=300\mathrm{\,k}$ compared with the corresponding monomer. (b) Comparison of extinction spectra for a nanowire dimer of intrinsic InSb with radius $R=100\mathrm{\,nm}$ and separated by a gap distance of $d=10\mathrm{\,nm}$ in vacuum at $T=200\mathrm{\,K}$, $300\mathrm{\,K}$, $350\mathrm{\,K}$ and $400\mathrm{\,K}$. Same material parameters as for the monomer in Fig.~3.}
\label{Fig:varygapandTemp}
\end{figure}
Similar to Fig.~3(a) for monomers, below the optical LSP resonance, the spectra in Fig.~6(a) show just one of the acoustic peaks, which is the acoustic LSP peak, while the others are invisible mainly because of the low mobility of the holes. As expected, the dimer enhances several resonances and also shows some new ones compared to the monomer. Nonlocal features are more pronounced for smaller gap sizes, and present for larger gaps than for metals. Most important for the present study is that the elusive acoustic resonance is also enhanced [compare the green curve with the others at low frequency in Fig. 6(a)]. This means that for the experimental observation of acoustic plasmon resonances it would be advantageous to consider dimer structures with small gaps.

 In Fig.~6(b) we see that higher temperatures shift resonances towards the blue, as expected, but also that the acoustic resonance at all temperatures considered is enhanced by the dimer as compared to the monomer. At higher temperatures the acoustic resonance is somewhat flatter and unpronounced, so it appears that increasing the temperature is not as advantageous for observing the acousic plasmon as decreasing the gap size would be [see Fig.~6(a)]. 

\section{Summary and conclusions}
In this paper, the two-fluid hydrodynamic Drude model was studied analytically and for the first time also numerically, allowing dimers to be studied. The model is an extension of the traditional, single-fluid HDM, and it accounts for two different kinds of free charge carriers (or fluids) such as e.g. electrons and holes, or light and heavy holes. 
We analyzed the two-fluid model for semiconductor nanowires and we focused on intrinsic InSb, although the model is expected to apply to other small-band-gap thermally excited semiconductors as well. 
 
For a low-loss single nanowire within the two-fluid model, many peaks other than the optical LSP are visible in the spectrum, also at frequencies below the optical LSP. None of these monomer resonances exist in the local-response Drude model, and several are not present in the single-fluid HDM either.
The additional resonances were identified using the dispersion relation of bulk systems within the two-fluid model. The model predicts two different longitudinal modes, optical and acoustic, where the latter exists as a propagating mode even down to zero frequency. An acoustic LSP and several acoustic bulk plasmon peaks were identified in the extinction spectra for single nanowires, all of which are characteristic of the two-fluid model. 

We used the analytical results for single nanowires to benchmark our COMSOL numerical implementation of the two-fluid model, and the quantitative agreement was excellent. This allowed us to continue with the numerical studies of dimers, where we focused on the acoustic modes. We showed that the dimer enhances the acoustic resonance in the extinction spectrum and also increases the field intensity in the narrow gap between the two nanocylinders.
The InSb intrinsic semiconductor nanowires with thermally excited charge carriers that we studied have just one visible acoustic resonance in their extinction spectra, namely the acoustic localized surface plasmon, while acoustic bulk peaks are invisible both for monomers and dimers. 
More importantly, for dimers we find that by decreasing the gap distance, nonlocal effects are increased and the acoustic resonance becomes stronger. This illustrates that, more generally,  it can be useful to design  semiconductor structures to increase the chances of experimentally observing acoustic plasmons. 

\section{Appendix: Linear equations for a single nanowire in the two-fluid model}
When applying the boundary conditions that are mentioned in the main text to the electric and magnetic fields, we obtain the following system of linear equations %
\begin{subequations}\label{Appendix}
\begin{equation}
\sqrt \ep_D J_n(x_D) + a_n \sqrt\ep_D H_n^{(1)}(x_D) = g_n\sqrt \ep_T J_n(x_T), 
\end{equation}
\vspace{-0.5cm}
\begin{eqnarray}
& J_n^\prime(x_D)+a_n H_n^{(1)\prime}(x_D)= \nonumber\\ 
 &g_n J_n^\prime (x_T) -\frac{in}{x1} h_{1n}J_n(x_1)-\frac{in}{x2} h_{2n} J_n(x_2), 
\end{eqnarray}
\vspace{-0.5cm}
\begin{eqnarray}
g_n\frac{in}{x_T} J_n(x_T) &+& h_{1n} \left(1+\frac{\beta_a^2\ep_{\infty} k_{L,1}^2}{\w_a^2(1+\alpha_1)}\right) J_n^\prime (x_1) \\
&+& h_{2n} \left(1+\frac{\beta_a^2\ep_{\infty} k_{L,2}^2}{\w_a^2(1+\alpha_2)}\right) J_n^\prime (x_2)=0, \nonumber
\end{eqnarray}
\newpage
\begin{eqnarray}
g_n\frac{in}{x_T} J_n(x_T) &+& h_{1n} \left(1+\frac{\beta_b^2\ep_{\infty} k_{L,1}^2}{\w_b^2(1+\alpha_1^{-1})}\right) J_n^\prime (x_1) \\
&+& h_{2n} \left(1+\frac{\beta_b^2\ep_{\infty} k_{L,2}^2}{\w_b^2(1+\alpha_2^{-1})}\right) J_n^\prime (x_2)=0. \nonumber
\end{eqnarray}
\end{subequations}
By solving this system of linear equations, the unknown coefficients $a_n$ can be found directly.

\section{Funding Information}
\noindent T.G. would like to thank the Ministry of Science, Technology and Innovation of Iran for the partial funds provided for her to stay and research at DTU. We acknowledge support from the Danish Council for Independent Research (DFF Grant No. 1323-00087). The Center for Nanostructured Graphene is sponsored by the Danish National Research Foundation (Project No. DNRF103). 
\section*{Acknowledgments}
\noindent We thank Emil C. Andr{\'e},  S{\o}ren Raza, Nicolas Stenger, and Sanshui Xiao for stimulating discussions.

\bibliography{Golestanizadeh_2019_nourl}


\end{document}